# Dr. Tux: A Question Answering System for Ubuntu users

**Bijil Abraham Philip**
University of Southern California
Los Angeles, CA 90007

bphilip@usc.edu

**Manas Jog**
University of Southern California
Los Angeles, CA 90007

mjog@usc.edu

**Apurv Milind Upasani**
University of Southern California
Los Angeles, CA 90007

aupasani@usc.edu

## Abstract

Various forums and question answering (Q&A) sites are available online that allow Ubuntu users to find results similar to their queries. However, searching for a result is often time consuming as it requires the user to find a specific problem instance relevant to his/her query from a large set of questions. In this paper, we present an automated question answering system for Ubuntu users called Dr. Tux that is designed to answer user's queries by selecting the most similar question from an online database. The prototype was implemented in Python and uses NLTK and CoreNLP tools for Natural Language Processing. The data for the prototype was taken from the AskUbuntu website which contains about 150k questions. The results obtained from the manual evaluation of the prototype were promising while also presenting some interesting opportunities for improvement.

## 1 Introduction

A Question Answering (QA) system aims at providing an answer to a user's question. Unlike Information Retrieval (IR) techniques, which return a set a documents that might contain the answer to the user's query, a question answering system has to return the answer itself. IR techniques can be used as first step in a QA system, to obtain a list of documents relevant to the question. Natural Language Processing (NLP) techniques can then be used to determine the best answer.

Most of the existing QA systems focus on answering general knowledge questions or questions that come up in a regular conversation. Automated technical QA is one domain where QA systems have immense potential that is yet to be tapped in to. Presently, if a user has a doubt related to use of a software or device or programming, the user has to depend on forums dedicate to the same. Forums like Stack Overflow and Ask Ubuntu provide a platform for users of differing levels of expertise to get their questions or doubts answered by their peers.

Using a search engine to find a solution to a technical question most often lead to links to threads in popular forums. Searching through different threads across different forums for the best solution is time consuming. Additionally, the forums are well moderated. If a user posts a question, similar to a question that has already been answered, the user is likely to get downgraded and down-voted.

We propose an automated technical QA system called Dr. Tux that utilizes questions and the best answers from technical forums. The objective of building such a system is to harness both IR and NLP techniques to answer technical question accurately and in reasonable time. Such system would enable users to find a solution to their query with reading through different threads across different forums. For pilot phase of this project we have restricted the domain to question related to Ubuntu.

There are numerous challenges associated with building a technical QA system. There is paucity of training data available for the technical domain. Standard QA techniques are not enough to tackle technical questions. For instance, Ubuntu, which is an Operating System is wrongly classified as a location by NER. Technical words are often classified as nouns and importance of the word is not captured. Our system relies on forums for questions and answers. However there is a difference between the way questions are asked in forums and the way questions are asked to QA systems. The questions asked in forums often contain a lengthy description of the problem, with along with code snippets and error logs. The questions posed to a QA system are more direct.

## 2   Related work

The idea of an automated QA system is quite old, but work in the field is still ongoing with many unsolved challenges remaining. Most of the older QA systems concentrated on a closed domain, while recent works have focused on building open domain QA systems for the web.

The TREC QA Track described in (Dang et al, 2007) focuses on open domain QA that directly returns the answer rather than a list of documents, in response to a natural language question. Since its inception, TREC has gradually evolved and can now answer both, factoid questions (i.e. questions based on facts) and list/ definition questions.

The QA system described in (Kuchmann-Beauger, and Aufaure, 2011) is based on a data warehouse. It provides composite answers composed of a dataset and corresponding chart visualizations. It uses surface patterns that incorporate business semantics and domain specific knowledge for question translation, thereby facilitating better coverage of questions.

QANUS, proposed in (Ng and Kan, 2010) is a publicly available, generic software framework for QA systems. The framework eliminates redundancy of code by implementing much of the code that are repeated across QA systems. The paper also presently a fully functioning factoid QA system built on QANUS called QA-SYS.

(Radev and Prager, 2000) describe a system to rank suspected answers to natural language questions. The proposed method uses predictive annotation method to process both query and the corpus. Predictive annotation augments phrases in texts with labels, by anticipating that they are targets for certain kinds of questions. For a given natural language question, an IR system returns a set of matching passages, which are then ranked using a linear function of seven predictor variables. (Mitamura et al, 2008) presents an overview of Advanced Cross-lingual information Access (ACLIA) task cluster. It provides a description of the complex question types that were evaluated, the metrics used for evaluating participant runs and the tools used to develop evaluation topics. It also provides details about the results of evaluating the submitted runs with the chosen metrics.

## 3   Data

For this project, we have used data from AskUbuntu which is a community driven Q&A website dedicated to the Ubuntu operating system. It is a part of the Stack Exchange Network. It provides users with a platform to ask questions and answer other user's questions related to Ubuntu. The websites are well moderated.

The reason we used we used data from Ask Ubuntu is because the posts, questions and answers are well structured, consistent and free of malicious and irrelevant answers. Their elaborate system of moderators ensures this. All questions are tagged with the relevant tags to indicate the topic being discussed. Every question has a short question title and a description. For every question, the questioner has the right to accept or reject answers posted by other users. Users also have the right to up-vote or down-vote an answer.

For this project, we downloaded a data dump from Ask Ubuntu, containing more than 153,000 questions and the related posts. The questions, associated tags and the accepted answers were then extracted from the dump file. In order to create a synonym set, a list of tags related to Ubuntu and their synonym tags was obtained using Stack Data Explorer. The Stack Exchange Data Explorer is an interface to query data from Stack Exchange. A set of duplicate questions and the original question they were related to, was obtained using the Data Explorer.

The data for testing the project was obtained from ten users, with varying level of expertise in Ubuntu. Each user was asked to fill up five questions related to Ubuntu and links to most similar question on Ask Ubuntu.

## 4 Architecture

Dr. Tux contains two layers: a natural language processing (NLP) layer and the Graphical User Interface (GUI) layer. The NLP layer is further divided into two components: Information Retrieval (IR) component and semantic similarity component. The Graphical User Interface (GUI) layer communicates with the NLP layer for providing the best answer for a user query.

The GUI is responsible for getting user queries in natural English and also for displaying the question that is most similar to the query along with its best answer. For retrieving the best question for a query, the GUI forwards the user input directly into the NLP layer which then returns the best matching question-answer set.

The NLP layer receives user input from GUI layer and first gives it to the IR component. The IR component performs POS tagging (using NLTK POS tagger) on the query followed by removal of stop words as well as words with unimportant POS tags. The words that remain are the keywords that are used in tf-idf based ranking of all questions in our dataset. We use the Gensim Python Framework for performing tf-idf analysis. For the purposes of speed, the NLP layer extracts only the top 20 questions from the tf-idf results. These questions are then redirected to the semantic similarity component.

In semantic similarity component, we perform semantic analysis on each result by comparing its semantics with the user query. Stanford CoreNLP was used in this component for generation of dependency trees that were then used for the semantic analysis. Once the similarity scores were calculated for each result, the results were then sorted in decreasing order with the most relevant question being on top. To ensure that the component does not return an empty answer we added a Python code that moves towards the next best question if no answer exists for the current best question. Finally the question and answer set is returned by the NLP layer to the GUI layer which then displays the answer to the user.

## 5 IR component

Performing semantic analysis over all 153k questions every time a user query is received will reduce the performance of the system dramatically. Since semantic analysis is a very expensive operation, we need to ensure that it is performed only for a specific subset of questions. The IR component is responsible for extracting question subset from the AskUbuntu corpus.

We used the Gensim Python framework for retrieving this question subset. The following were the algorithms that were tried from the Gensim module during the experiment: tf-idf, Latent Symantic Analysis (LSA) and Latent Dirichlet Allocation (LDA). Both LSA and LDA were extremely memory intensive making it very difficult to execute them even on machines with 4GB of RAM. An attempt was made to check if it was possible to use them by only using question titles from AskUbuntu. The results were unsatisfactory since most of the question titles contained information that was not enough to guess the question. The tf-idf algorithm, however, performed much better and its performance was also substantially better than the other algorithms that were tried upon and hence tf-idf was finally selected for usage in our IR component.

The question title and the question body were extracted from the dataset separately and hence there was an issue about what combination should be used during the tf-idf analysis. Treating question body and question title as a single document gave worse results than using only one of them. However, the weighted average of tf-idf scores calculated for the question title and question body separately performed much better. For the prototype, we assigned equal weights to the tf-idf scores calculated.

## 6 Semantic similarity component

The Semantic similarity component uses the dependency parser from the Stanford CoreNLP for calculation of similarity scores. These scores are calculated by trying to find out the semantics that are common between the user query and the question set obtained from the IR component.

The first step is to first find out the type of question asked. In this system, the questions are grouped into 2 categories:

- Seeking general information about something (Factual).
- Trying to find out the reason for a software problem (Troubleshooting).

The reason for grouping questions into categories mentioned above is because the troubleshooting queries require a higher amount of semantic analysis when compared to factual queries. It was also observed that the troubleshooting question does not describe the concept. This is obvious because it is assumed that the user was aware about a concept and that circumstances other than the knowledge of the concept have led to this problem. This is very different from querying a factual question, since the user is unaware of the concept used. Therefore if the user intends to know about a concept, then we consider only factual questions from the result, otherwise we consider troubleshooting questions.

The classification approach can be used for grouping the question subset. Due to the lack of annotated data, it was assumed that if the user is asking a troubleshooting question, then the sentence will contain either a negation of a verb or a verb with negative sense. Dependency parser can be used for detecting negation in a sentence while the negative list of words can be generated by performing sentiment analysis on some training data. Again, because of the lack of training data, the negative opinion words were taken from the list that was generated by opinion analysis on web data.

The classification module is used for considering only those questions from the top 20 questions that belong to the same category as the user's query. These questions will be referred to as candidate questions from here on.

The final phase of semantic analysis involves knowing the relative importance of the words. Since this system deals with the Ubuntu users, we only need to consider words that are related to Ubuntu. The list of such words was obtained by downloading all the tags that are used in the AskUbuntu website. The tags include names of various OS versions, processes and applications related to Ubuntu and also other terminology that specialized in the Ubuntu user domain. The new words that were added into our system are not available in Wordnet and hence there was a need to create a synonym set customized for this domain. The list of tag synonyms that had been generated by the AskUbuntu community was used as a solution for the above problem.

The tags downloaded were used for generating the word vector for the query as well for the candidate questions. For generating the vector for a question, the system first considers the root word from its corresponding dependency tree. Then it identifies all the tags and the distance of those tags from the root word. This distance is taken as the number of words between the root word and the tag. The list of all tag-distance pairs are used for creating the word vector. The idea behind using such a method is that the root word represents the theme of the sentence and the closer a tag is to that theme word, the more relevant it is to the theme of the sentence. For example, let's consider the question:

```
How do I install Ubuntu on Windows?
```

We know that Ubuntu and Windows are computer related terms and the user is talking about installation, specifically installation of Ubuntu on Windows and not the other way around. The tag list used in the system contains these terms and hence it can be used for finding out those terms. The dependency tree returns *install* as the root word and therefore the word vector is:

$$\{Ubuntu: 1, Windows: 3\}$$

Clearly, the word vector was able to retain the information inferred from the sentence. The component finds out the word the word vector for the query and all the candidate questions. The cosine similarity for query's word vector Q and a candidate's word vector C is calculated as:

$$\cos\theta = \frac{Q \cdot C}{\|Q\|\|C\|}$$

The final similarity score is calculated by the following formula:

$$similarity = \cos\theta \times (\text{tf-idf score})$$

Similar operation is performed for other candidate questions and the candidate questions are then arranged in the descending order of their similarity scores.

Finally the component checks if the top candidate question contains a non-empty answer. If answer exists, the result is returned to the GUI component, otherwise we move on to the next best question.

## 7. Evaluation

One of the key areas of this project involves the evaluation of the system. Unlike other NLP systems however, it is very difficult to evaluate a Question Answering system, much less for a technical system. Since the questions are of technical nature, the satisfaction measure of the answer is highly subjective. For this project, we had initially decided to score the responses as following:

| Criteria | Measure |
|---|---|
| Unacceptable | 0 |
| Acceptable | 1 |

However, a careful failure analysis (covered in section 6) revealed that in many cases, the answer generated by the system was either too specific for a question or partially correct. Eg. When asked about laptop overheating, it would give the answer for laptop overheating for Acer Laptops. This answer in a way was partially correct and in some cases may help in overcoming the overheating problem of many laptops. Therefore, it would be wrong to discard the answer all together.

To resolve this, we developed a new metric for calculation where we scored the responses into 3 criteria:

| Criteria | Measure |
|---|---|
| Unacceptable | 0 |
| Fair | 1 |
| Acceptable | 2 |

Our first method of evaluation was a manual method. We asked 10 users ranging from expert users to naïve ubuntu users to ask 5 questions each. We ensured that the questions had no duplicates and manually fed the questions to the system. The responses were logged and users were asked to score the responses. These scores were then added to form an overall score.

The first iteration of evaluation yielded a score of 51/100 with the breakdown of score as follows:

| Criteria | Questions | Score |
|---|---|---|
| Unacceptable | 14 | 0 |
| Fair | 21 | 21 |
| Acceptable | 15 | 30 |
| Total | 50 | 51 |

We did an architectural analysis (section 6) and employed a new mechanism for question comparison which resulted the following results in second iteration.

| Criteria | Questions | Score |
|---|---|---|
| Unacceptable | 9 | 0 |
| Fair | 11 | 11 |
| Acceptable | 30 | 60 |
| Total | 50 | 71 |

Nyberg et.al mentions several criteria for evaluation of QA systems. We plan to employ few of these techniques as a part of our evaluation criteria.

**User feedback**

1) Direct approach

   User will be randomly asked to evaluate the responses. The user's evaluation will be logged for future analysis.

2) Indirect approach

   The questions asked by the user will be logged and those questions will be evaluated with their responses. The intuition behind this approach is that the user's will ask the variations of same questions in case of unsatisfactory responses from the system.

Apart from this, we have already performed one set of architectural evaluation. In this, we have tested every module, keeping the accuracy of the module in context. Section 6 will describe the architectural analysis results in detail. We plan to perform the architectural evaluation iteratively in following cases:

1) When new data from AskUbuntu is added into the system.
2) When data from similar domain , like Ubuntu forums is added.
3) When we try to integrate the cross domain data from systems like StackExchange and StackOverflow.

## 8. System Analysis

### 8.1 NLP component analysis

The analysis of NLP component can be further divided into 2 components:

#### 8.1.1 IR component analysis

For evaluation of IR component, we scraped UbuntuForums and downloaded a set of 423 questions related to installation and troubleshooting of Ubuntu. We applied these questions to our IR component to evaluate whether the component provides high recall. We observed that for 376 questions, the most relevant answers are obtained in the top 20 responses of the Gensim TF-IDF framework. The accuracy of the IR component was hence calculated to be around 89%. It is important to note that the scraped questions were from a different domain all together and they were not moderated (unlike AskUbuntu).

The reason for choosing 20 top responses was due to the fact that after careful analysis of responses, we observed that later responses yielded a very low TF-IDF score, rendering them of a very little use for semantic matching.

#### 8.1.2 Semantic similarity component analysis

Semantic similarity component primarily does a classification of a query into factual and troubleshooting queries. We use a combination of dependency parsing and negative word list to find whether the query is factual or troubleshooting. We find that though this approach works extremely well for the factual queries, it sometimes gives mixed results for troubleshooting queries. In fact, most of the queries that are successful or work fairly are factual queries. Failure analysis mentions in detail the failure reasons.

### 8.2 Dialogue component analysis

Dialogue component currently consists of a salutations component which takes care of simple user interaction like Hi, Hello and Bye. The UI component also provides ability for system to work in debug mode which prints out necessary logs for debugging. Here we primarily analyze the response time taken for the query to process.

For 50 queries, we observe that the average response time was less than 10 seconds.

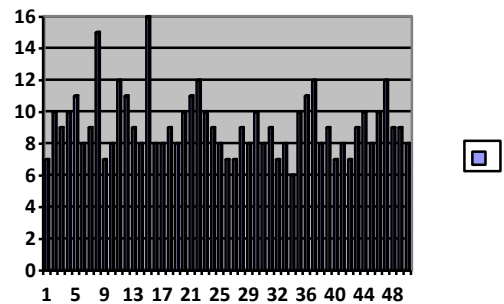

### 8.3 Comparison with baseline systems

If we compare the system with a normal QA system, this system at the current stage gives less accuracy. This is mainly due to two reasons.

Firstly, there is dearth of training data. Most of the data on the forums is not moderated properly and maybe hugely error prone. The data present may be addressing a specific problem or it may not be present. This makes it difficult for us to handle complex factual queries.

Secondly, unlike most QA systems, since our domain is technical, we have troubleshooting queries, for which we have to perform a distance measure from the root to form a word vector. This is because it is extremely difficult to find the subject on which the query is based. It is one of the novel features which are present in our system.

### 8.4 Comparison with comparable systems

Unlike most QA systems which base their knowledge base on factual questions, our system is responsible for assisting the user in his technical needs. We have yet to see a QA system which handles technical domain. Plus our system is scalable and can adapt to multiple domains. It performs resolution of factual queries with a high accuracy i.e. we were able to observe that for most factual queries, the system was able to generate at least a fair response. However, our system can handle troubleshooting queries to a certain extent too.

### 8.5 Failure analysis

For a set of 50 questions, we observed that 38 questions were factual while 12 were troubleshooting questions. Following is the table which shows the results per type.

| Criteria | Factual | Troubleshooting |
|---|---|---|
| Unacceptable | 2 | 7 |
| Fair | 9 | 2 |
| Acceptable | 27 | 3 |

#### 8.5.1 Data Unavailability

One of the main reasons for unacceptable responses in factual data was the lack of relevant data in the set of questions. Furthermore there are several cases, where the system gives a fair response because the most matched question lacked an answer, thereby prompting the system to choose the question with lesser similarity measure. This problem propagates to some extent even for troubleshooting questions.

#### 8.5.2 Inability to pinpoint the problem source

This is the major problem we face with troubleshooting questions. Since we use the distance from the root as measure for finding the problem source, it is highly dependent on parser output and sometimes pinpoints wrong problem source. Eg. My Ubuntu does not boot when installed with Windows represents Windows as the problem source whereas the problem source is Ubuntu. This results in imperfect word vector mapping which often result in inconsistent outputs.

#### 8.5.3 Improper classification

Currently, we classify the factual and troubleshooting queries based on the NEG tag in dependency parser output or word present in list of negative words. There have been cases wherein the system has wrongly classified a factual question as troubleshooting one and vice versa. This has resulted in improper classification and the document has not been chosen for further processing. We need a better mechanism for classification of queries.

## 9. Future Directions

We plan to incorporate several new features time. Firstly we would like to incorporate an error detection and correction mechanism for technical domain. It will greatly reduce the syntactic and semantic errors that can be used to clean both the incoming data as well as the user query.

Secondly, we plan to extend our database to similar forums like UbuntuForums . However we need to consider the difference in question/answer quality as the questions may not be moderated.

Thirdly, we want to extend our approach to resolve programming or math questions by obtaining data from sites like StackOverflow. It is one of the key challenges that we look forward to as it would incorporate a lot of pattern and code analysis for fetching the responses.

Finally, we would like to develop a dialogue system for higher interactive purpose. It would be very useful for the user, especially for resolution of troubleshooting questions.

**Appendix:**

### Individual Contributions

| Module | Approach used | Contributer |
|---|---|---|
| Data collection | Yes | Apurv, Bijil, Manas |
| IR tf-idf module | Yes | Manas |
| Question semantic similarity analyzer module | Yes | Bijil , Manas |
| Improving IR using wordnet | No | Apurv |
| GUI and simple dialog component | Yes | Apurv - Bijil |
| Semantic parsing and Ubuntu synset cration | Yes | Bijil |
| Manual evaluation | Yes | Apurv, Manas |
| Test data generation for IR | Yes | Apurv |
| Collection of data from Ubuntu users | Yes | Bijil |
| Failure Analysis | Yes | Apurv – Bijil |
| Cosine similarity analysis | Yes | Manas |
| GUI component analysis | Yes | Apurv-Bijil-Manas |
| Module | Approach used | Contributer |
| Data collection | Yes | Apurv, Bijil, Manas |
| IR tf-idf module | Yes | Manas |
| Question semantic similarity analyzer module | Yes | Bijil , Manas |